\documentstyle[aps, preprint]{revtex}
\begin{document}
\title{Finite size effects in
metallic superlattice systems}
\author{J.~Chen}
\address{Physics Department and Winnipeg Institute
for Theoretical Physics,\\
University of Manitoba,
Winnipeg, Manitoba R3T 2N2, Canada}
\author{R.~Kobes}
\address{Physics Department and Winnipeg Institute
for Theoretical Physics,\\
University of Winnipeg,
Winnipeg, Manitoba R3B 2E9, Canada}
\author{J.~Wang}
\address{Physics Department and Winnipeg Institute
for Theoretical Physics,\\
University of Manitoba,
Winnipeg, Manitoba R3T 2N2, Canada}
\maketitle
\begin{abstract}
Clean metallic superlattice systems composed of alternating layers
of superconducting and normal materials are considered, particularly
aspects of the proximity effect as it affects the
critical temperature.
A simple model is used to address the question
of when a finite--sized system theoretically
approximates well a true infinite
superlattice. The methods used in the analysis afford
some tests of the approximation used that the
pair amplitude of the Cooper pairs is constant over a
superconducting region. We also use these methods to construct
a model of a single superconducting layer which intends
to incorporate a more realistic form of the pair amplitude
than a simple constant.
\end{abstract}
\section{Introduction}
Metallic superlattice systems consisting of alternating layers of
superconducting and normal films have attracted wide interest, both
experimentally and theoretically \cite{review}. Much of the theoretical
work has gone into the understanding of the proximity effect in such
systems. This effect between adjacent layers of materials with different
superconducting properties has been studied extensively for the case
of two such neighbouring films. Close to the transition point
Ginzburg--Landau theory can be employed [2--7], but away from this
point other approaches such as the constant pair amplitude
approximation [8--15], a ``Cooper'' proximity effect argument \cite{cooper},
the tunneling model of McMillan \cite{tunnel}, a WKB approximation
\cite{wkb}, or a semi--classical approximation \cite{semi} can be used.
\par
Implicit in these approaches when applied to superlattices
is the argument that, from symmetry
considerations, the form of the pair amplitude of the Cooper
pairs for a bilayer is the same as that for a
superlattice consisting of an infinite number of adjoining bilayers.
While true for an infinite superlattice,
experimentally these systems are of a finite extent, and one could
then question the reasonableness of the bilayer assumption for
finite systems. This issue could also arise
in the general
context of the applicability of Bloch's theorem for periodic
potentials in cases of large but finite--sized systems, as for
instance as applied to the Kronig--Penney model \cite{kittel}. In this
paper we will address some aspects of
this question of how many bilayers are effectively needed to have
an infinite superlattice system, particularly as related to
how the critical temperature of such systems is affected.
\par
The paper is organized as follows. To introduce the approach
and to establish notation, in Sec.~II we review a method of analyzing the
proximity effect based on the constant pair amplitude
approximation as applied to the simple case of a single
superconducting film immersed in a normal metal \cite{white}.
In Sec.~III we extend this method to an infinite superlattice
composed of alternating superconducting and normal layers. To test some
of the ideas we shall later employ, we consider
the superlattice as the limit of a finite sized system as the
number of layers approaches infinity; the result we find for the
critical temperature agrees with that obtained by arguments
which explicitly incorporate Bloch's theorem using the same general
approach \cite{yuan}. In Sec.~IV we consider finite--sized ``superlattices''.
We argue from the calculated form of the pair amplitude
that the approximation
of a constant pair potential is reasonable in two distinct
cases: for thick superconducting
films over a wide range of thicknesses of the normal film,
and for thin superconducting and normal films. As related to
finding the critical temperature in this approach, in the first case
a relatively small number of layers are needed
before the finite--sized system approximates well
an infinite superlattice, whereas in the second case
quantitative differences arise between the finite and infinite lattices.
In Sec.~V we reconsider the single superconducting
layer of Sec.~II regarding the issue raised in the previous section
of the reasonableness of the constant pair potential assumption.
We study a model based on these methods which aims
to incorporate in a simple manner a more realistic form of the pair
potential, and find that the resulting zero temperature
critical thickness
gets enhanced compared to that of Sec.~II found using the constant pair
potential approximation. Sec.~VI contains some concluding remarks.
\section{A Single Film}
We first study the case of a single superconducting film
embedded in a normal metal \cite{white}. The film, of width $2a$, is
centered at the origin $x=0$, as in Fig.~1.
The Hamiltonian for this system in terms of the electron
wave function $\psi$ is given in the mean field approximation by
\begin{eqnarray}
 H& = & \int d^3r\,g(x) \mid F(x)
\mid^2+\int d^3r\, \phi^{\dag}({\bf r})
\left[-{\nabla^2 \over 2m}-\mu\right]\tau_3\phi({\bf r})
  \nonumber\\ & - & \int d^3r\,g(x) F(x)
\phi^{\dag}({\bf r})\tau_1\phi({\bf r}),
\label{singleham} \end{eqnarray}
where $\tau_i$ are the Pauli matrices, $F(x)=F^*(x)=-\langle
\psi_{\uparrow}(x)\psi_{\downarrow}(x)\rangle$ is the pair
amplitude,
\begin{equation}
\phi({\bf r})= \left\{ \begin{array}{c}
\psi_{\uparrow}({\bf r}) \\
\psi_{\downarrow}^{\dag}({\bf r}) \end{array} \right\},
\end{equation}
and $g(x)$ denotes the spatially inhomogeneous
BCS coupling constant.
We assume the superconducting and normal metals are clean and
have the same effective electron masses, Fermi velocities, Debye
temperatures, etc.
\par
The Nambu doublet is then decomposed as
\begin{eqnarray}
 \phi({\bf r},t)& = &\sum_i {1 \over (2 \pi)^3}\int dE\int
d^2l\,[u^{(i)}(E,l,x)a^{(i)}(E,l)e^{-i(Et-
     {\bf l} \cdot{\bf \rho})} \nonumber\\          & + &
v^{(i)}(E,l,x)b^{(i)\, \dag}(E,l)e^{i(Et-
    {\bf l} \cdot{\bf \rho})} ],
\end{eqnarray}
where  ${\bf l}=(k_y, k_z)$,
${\bf\rho}=(y,z)$, and ``$i$'' labels the particular
linearly independent solution present. If we now introduce
\begin{equation}
u^{(i)}(E,l,x)=\chi^{(i)}(E,l,x)e^{iqx},
\end{equation}
where $q=\sqrt{k_F^2-{\bf l}^2}$, then assuming
$\chi(E,l,x)$ to be smooth on the atomic scale results
in the Bogoliubov equation following from Eq.(\ref{singleham}):
\begin{equation}
\left\{E+i{\tau_3q \over m}{d \over
 dx}+g(x)F(x)\tau_1\right\}\chi^{(i)}(E,l,x)=0.
\label{bogeq}\end{equation}
The corresponding wave function $v(E,l,x)=u(-E,l,x)$ can then
be readily obtained.
\par
The non--linearity introduced by
$F(x)=-\langle
\psi_{\uparrow}(x)\psi_{\downarrow}(x)\rangle$
makes Eq.(\ref{bogeq}) difficult to solve for inhomogeneous
geometries. To obtain a simple solution we therefore
make the approximation
\begin{equation}
 g(x)F(x)=\left\{ \begin{array}
{ll}\Delta & \mbox{for $-a<x<a$}           \\
0 & \mbox{for $x<-a$ or $x>a$} \end{array} \right. ,
\end{equation}
where $\Delta$ is a constant. The solutions to Eq.(\ref{bogeq})
under this approximation are readily found to be
\begin{equation}
 \chi(E,l,x)= \left\{
\begin{array}{ll} A\left(
\begin{array}{cc}      1 \\
0 \end{array} \right) e^{ikx} \ , \
B      \left( \begin{array}{cc}
     0 \\ 1 \end{array} \right)
e^{-ikx}       & \mbox{for $x<-a$ or $x>a$},
 \\      C \left(
\begin{array}{cc}      {1+\gamma} \\
\delta \end{array} \right)
e^{ipx} \ , \  D \left(
 \begin{array}{cc} \delta \\
{1+\gamma}      \end{array}
\right)e^{-ipx}
 & \mbox{for $-a<x<a$}
\\
   \end{array} \right. ,
\label{solutions}\end{equation}
where
\begin{eqnarray}
& &p=\gamma k =mE \gamma/q = mE\gamma/\sqrt{k_F^2-{\bf l}^2}, \nonumber\\
& &\delta=-\Delta /E, \nonumber\\
& &\gamma=\sqrt{1-\delta^2}.
\end{eqnarray}
The coefficients of the wave function entering into the
solutions of Eq.(\ref{solutions}) can be determined
by imposing continuity at the interface $x=\pm a$ together
with the normalization conditions
\begin{eqnarray}
& &\int_{-\infty}^{\infty}dx\,{\overline u}(E,l,x)u(E',l,x)
=2\pi\delta(E-E')=
\int_{-\infty}^{\infty}dx\,{\overline v}(E,l,x)v(E',l,x),\nonumber\\
& & \int_{-\infty}^{\infty}dx\,{\overline u}
(E,l,x)v(E',l,x)=0.
\end{eqnarray}
\par
Thus far these solutions contain the free parameter $\Delta$.
We fix this parameter by requiring that it minimize the free
energy of the system:
\begin{equation}
\left\langle \frac{\partial H}{\partial\Delta}\right\rangle=0,
\end{equation}
where the angular brackets denote the thermal average.
Using Eq.(\ref{singleham}) we find that this condition
translates into the self--consistent gap equation
\begin{equation}
\Delta = {g  \over 2a} \int_{-a}^{a} F(x, \Delta)\, dx.
\label{gap}\end{equation}
In other words, we demand that the spatial average of the
calculated pair amplitude be equal to the constant $\Delta$
assumed in finding the solution.
\par
In the limit $\Delta\to 0$ the gap equation (\ref{gap}) will
determine the critical temperature $T_c(a)$ as a function of
the thickness $a$ of the superconducting film. Using the
solutions and boundary conditions discussed previously, we
find this equation becomes, assuming $T_c(a)\ll T_d$,
\begin{equation}
\frac{1}{gN(0)}=
H(\omega_d)-{\beta \over 2}\int_{0}^{\infty}d \omega \,
H(\omega)\cosh^{-2}({\beta \omega / 2}) ,
\label{critsingtemp}\end{equation}
where $\beta=1/k_BT$, $N(0)=mk_F/2\pi^2$ is the
density of states of one spin projection
at the Fermi surface and,
 with $\Lambda=2k_Fa$ and $\Omega=\omega/E_F$, the function
$ H(\omega)=H(\Lambda \Omega)$ is given by
\begin{eqnarray}
& &H(x)=\gamma_e+\ln(x)-{1 \over 2}G(x),\nonumber\\
& &G(x)=\cos(x)-{\sin(x)\over x}
+2\,{\rm Ci}(x)+x\,{\rm si}(x).
\label{hofx}\end{eqnarray}
Here ${\rm Ci}(x)$ and ${\rm si}(x)$ are the Cosine and
 Sine integrals defined as \cite{as}
\begin{eqnarray}
& &{\rm Si}(x)={\rm si}(x)+\frac{\pi}{2}=
\int_{0}^{x}{\sin t \over t} dt,\nonumber\\
& & {\rm Ci}(x)=\gamma_e +\ln x +\int_{0}^{x}
{\cos t -1 \over t } dt,
\label{sincos}\end{eqnarray}
and $\gamma_e$ is Euler's constant. The function $H(x)$ of
Eq.(\ref{hofx}) has the following limiting forms:
\begin{eqnarray}
& &\lim_{x \rightarrow \infty}H(x)=
\gamma_e+\ln(x)+O(1/x),\nonumber\\
& &\lim_{x \rightarrow 0}H(x)={\pi x \over 4}+O(x^2).
\end{eqnarray}
\par
For comparison with other geometries the numerical solution
to Eq.(\ref{critsingtemp}) giving the critical temperature
$T_c(a)$ will be discussed in Sec.~IV. For now we only note that
Eq.(\ref{critsingtemp}) can be solved analytically in the zero
temperature limit, which gives the critical thickness $a_c$ below
which superconductivity cannot be maintained at any temperature:
$T_c(a_c)=0$. One finds, assuming $\Lambda_c\Omega_d\gg 1$,
\begin{equation}
a_c={1 \over \pi k_F}\left({T_F \over T_c}\right)
=\frac{1}{2}\pi e^{-\gamma_e} \xi_0
\approx 0.882\xi_0,
\end{equation}
where $T_c$ is the bulk critical temperature
and $\xi_0=v_F/\pi \Delta_0$ is the bulk coherence length.
This value of the critical thickness compares favourably
with that found using other approaches \cite{7,tunnel}, and will be used in
subsequent sections as a convenient length scale when discussing
other geometries. It is also useful for later comparisons with
other geometries to introduce $a_c$ directly into the
gap equation of Eq.(\ref{critsingtemp}); doing so
leads to the following equivalent forms:
\begin{eqnarray}
& &\ln\left[\frac{a}{a_c}\right]=I(\tau),\nonumber\\
& &\ln\left[\frac{T_c}{T_c(a)}\right]=\frac{\pi}{8\tau}+\frac{1}{2}K(\tau),
\label{critsing}\end{eqnarray}
where
\begin{eqnarray}
& &\tau=\frac{2}{\pi}\frac{a}{a_c}\frac{T_c(a)}{T_c},\nonumber\\
& &I(x)={1 \over 2}\pi x \ln 2-x\int^{\infty}_{0} {dt \over t^2}
{\sin(2t)\over 2t}\ln[1+\exp(-2t/x)],\nonumber\\
& &K(x)=\int_{1}^{\infty}dt\left[1-\frac{1}{t^2}\right]
\ln\tanh\left[\frac{\pi x t}{2}\right].
\label{tau}\end{eqnarray}
The feature of these and the following gap equations that the
thicknesses of the superconducting and normal layers scale with
$a_c$ makes it possible that dirty materials might
also be described within the present framework by a simple
reinterpretation of $a_c$, as was found in the single layer
case by a comparison with the tunneling model \cite{white,tunnel}.
\section{Infinite Superlattice}
We next study an infinite superlattice consisting of alternating
superconducting films of width $2a$ separated by
normal films of width $2b$. This is illustrated in Fig.~2.
Using the approach of the previous section, we first consider
$N$ such films and eventually take $N\to\infty$. We assume the
pair potential $g(x)F(x)$ in each superconducting layer is a constant
$\Delta$ and vanishes in each normal layer. This allows us to
solve the corresponding Bogoliubov equations (\ref{bogeq}) in
each region subject to the appropriate boundary conditions.
The parameter $\Delta$ is then determined by the minimization
of the free energy of the system. This condition, in the limit
$\Delta\to 0$, gives the critical temperature $T_c(a,b)$ as a
function of the thicknesses $a$ and $b$ of the films. Taking the
limit $N\to\infty$ then results in the gap equation
\begin{equation}
\ln(a/a_c)=I(\tau)-J(\tau,y),
\label{gapinf}\end{equation}
where $y\equiv 1+b/a$, $\tau$ and $I(x)$ appear in Eq.(\ref{tau}), and
\begin{equation}
J(x,y)=x\int_{0}^{\infty} {dt \over t}
\frac{\sin^2(t)}{t^2}\cot[ty]
\ln\cosh(t/x).
\end{equation}
\par
For infinite $b$ the function $J(\tau,y)$ vanishes,
resulting in the gap equation of Eq.(\ref{critsing}) for the
single layer. $J(\tau,y)$ thus represents coherence effects
the other
layers introduce to the single film geometry.
Eq.(\ref{gapinf}) agrees with the results of an analysis
directly incorporating Bloch's theorem in this context
for periodic systems \cite{yuan}. We present in the next section
the numerical solution to Eq.(\ref{gapinf}) in various
cases and compare these results with those corresponding to
a finite number of layers using this same general approach.
\section{Finite Superlattice}
In this section we examine by these methods
 a ``superlattice'' with a finite number of layers.
As with the true infinite superlattice this consists
 of alternating
superconducting films of width $2a$ separated by
normal films of width $2b$, as in Fig.~2, but now after
the last superconducting layer on each end
we assume a normal metal
fills the remainder of space. We again assume the
pair potential $g(x)F(x)$ in each superconducting layer is a constant,
but unlike the infinite superlattice there is no reason
{\it a priori} to assume the same constant in each layer.
We therefore assume
$g(x)F(x)$ is some constant $\Delta_j$ for the $j^{\rm th}$
superconducting layer, and as before
vanishes in each normal layer. This is illustrated in
Fig.~3. With appropriate boundary
conditions the Bogoliubov equations (\ref{bogeq}) are readily
solved, and the
parameters $\Delta_j$ are again
 determined by the minimization
of the free energy of the system. This again translates into
equating the spatial average of the calculated pair potential
in the $j^{\rm th}$ layer to the assumed constant $\Delta_j$.
These conditions now are a coupled set of
$N+1$ equations, which in the limit
$\Delta_j\to 0$ and $\Lambda\Omega_d\gg 1$ are given by
\begin{eqnarray}
& &\ln\left[\frac{T_c(a,b,N)}{T_c}\right]=-\frac{\pi}{16\tau}
\nonumber\\
& &-\frac{1}{4}\left\{K(\tau)-2jy
K(2jy\tau)+(1+2jy)K[(1+2jy)\tau]\right\}
\nonumber\\
& &+\frac{1}{4}\sum_{l=j+1}^{N}\frac{\Delta_l}{\Delta_j}
\left\{ 2(l+j)yK[(l+j)y\tau] + 2(l-j)yK[(l-j)y\tau]
\right.\nonumber\\
& &-[1+(l+j)y]K[1+(l+j)y\tau]
-[1+(l-j)y]K[1+(l-j)y\tau]\nonumber\\
& &\left.-[-1+(l+j)y]K[-1+(l+j)y\tau]
-[-1+(l-j)y]K[-1+(l-j)y\tau]\right\},
\label{critfin}\end{eqnarray}
where $j=0,1,2,\ldots,N$, $y=1+b/a$,
and $\tau$ and $K(x)$ appear in Eq.(\ref{tau}).
As a check, if we consider the single layer limit with $N=0$, then
the system of equations (\ref{critfin}) reduces to a single
equation, which agrees with the gap equation (\ref{critsing})
derived before.
\par
We first consider the $j=N^{th}$ equation of the set (\ref{critfin}),
for which the term involving the summation over $\Delta_l/\Delta_j$ does
not contribute. This equation will determine the critical
temperature $T_c(a,b,N)$ as a function of the thicknesses $a$ and
$b$ of the films and of the number of superconducting layers
present, $2N+1$. In Figs.~4 and 5 we present the numerical solution
to this equation for different values of $b$ and $N$, along
with the results for the infinite
superlattice of Eq.(\ref{gapinf}) and for the single
layer of Eq.(\ref{critsing}). Although for all values of $b$ and $N$
the correct bulk limit as $a\to\infty$ is reached, quantitative
differences with the infinite superlattice are obtained for
smaller values of $a/a_c$ for the finite lattice. In particular,
 as $b/a_c$ increases in the finite lattice the curves do not approach
that of the single layer case, as might be expected and is
found for the infinite superlattice.
\par
Before concluding anything from these results, however, let us
examine the other equations of the set (\ref{critfin}). For
$j=0,1,2,\ldots,N-1$ these equations involve the term containing
the summation over $\Delta_l/\Delta_j$, and can be used to
express, for example, $\Delta_1,\Delta_2,\ldots,\Delta_N$
in terms of the value of the center layer, $\Delta_0$. In
Figs.~6--9 we present the numerical solution to these equations
at zero temperature for some representative values of $b/a_c$
and $N$. We would expect that as one moved out from the central
region the superconductivity would become weaker in some sense,
which would translate into decreasing values of the ratio
$\Delta_j/\Delta_0$ as $j$ increases. This expectation is borne
out for smaller values of $b/a_c$ for a large range of $N$, but
not for larger values of $b/a_c$. At finite temperature,
illustrated in Figs.~10 and 11, the expected behaviour is again
found for smaller values of $b/a_c$ but not completely for the larger values,
although the situation improves as the temperature
increases. What one might
conclude from this analysis is that the method used in this context
seems reliable for thin normal films over a wide range of thicknesses
of the superconducting layers, but for thicker normal films
the method seems reasonable only for thicker superconducting
layers (i.~e., larger critical temperatures).
This could indicate that the approximation of a constant pair
potential in the superconducting layer appears reasonable
in this range of validity, for which little spatial variation in the true
potential might be expected.
\par
With the preceding discussion in mind we now return to a comparison
of the finite superlattice results of the critical
temperature in Figs.~4 and 5 with that of the infinite
superlattice. We restrict this comparison to a potentially
large but still finite number of layers in the finite system,
as in the limit $N\to\infty$ our initial assumption of the
independence of the gap parameters $\Delta_j$ in
each layer will no longer be valid. Given this,
for thicker superconducting layers ($a/a_c>10$),
where the approximation of a constant pair potential
seems reasonable over a wide range of thicknesses $b/a_c$ of the
normal layer, we could conclude for these methods
that a finite superlattice approximates well an infinite
superlattice for a reasonably small number of layers ($N\sim 10$).
On the other hand, for thinner superconducting layers ($a/a_c<10$),
the approximation of a constant pair potential
seems reasonable only for thin normal layers ($b/a_c<0.1$). In this
range we could conclude that there are noticeable differences
between the finite lattice
and the infinite superlattice in this approach, as seen from Fig.~4,
even with a relatively large number of layers for the finite system.
Qualitatively what is found is that, for a given thickness $a/a_c$,
the corresponding critical temperature
in the finite lattice is smaller compared to the case of the
infinite superlattice. This feature could be understood
in terms of the form of the pair potential suggested by Fig.~3
and found in Figs.~6, 7, 10, and 11,
where the gap parameters decrease relative to the central value
as one moves towards the edges. If one assumes this gap parameter
ratio measures in a way the relative strength of the binding of the
Cooper pairs in each region, then from this one might infer that
the superconductivity is in a sense ``weaker'' towards the
edges for the finite potential of Fig.~3
compared to the case of the same constant
gap parameter assumed over each superconducting region in an infinite
superlattice, as in Fig.~2. Thermal excitations would then
more readily break superconductivity in these outer regions,
leading to a lower critical temperature for a given thickness
$a/a_c$ in the finite lattice compared to the infinite superlattice.
Such a tendency of weaker superconductivity in the finite lattice
continues to hold except as the thickness of the
superconducting layer decreases
further ($a/a_c<0.1$); in the extreme limit this is
a reflection of the fact that the approximation
$\Lambda\Omega_d=2k_Fa\Omega_d\gg 1$ used in deriving
Eq.(\ref{critfin}) is breaking down.
\section{The Single Layer Revisited}
The results of the last section raised questions concerning
the appropriateness of the
assumption of a constant pair amplitude in a superconducting
layer. In this section we will attempt to improve on this
assumption for the case of the single layer described in
Sec.~II. Specifically, we imagine for the situation
of $2N+1$ layers in Fig.~2 letting the
width $b$ of the normal layer approach zero. In effect we have then
artificially divided a superconducting single layer
of width $2A$, as in Fig.~12, into
$2N+1$ layers, each of width $2a$.
As in the last section we next assume the gap
parameter in each layer is some constant $\Delta_j$, and then
set up and solve the system of self-consistent gap equations
found by a minimization of the free energy. This again leads
to a coupled set of equations, and as in the
previous section the $j=N^{\rm th}$ equation for the layer on the
extreme edge will determine the transition temperature as a
function of the film thickness $a$ and the number of layers $N$.
One obtains
\begin{eqnarray}
& &\ln\left({a \over a_c}\right) = {1\over 2}\left[2N\ln(2N)
-(1+2N)\ln(1+2N)\right]+\frac{1}{4}\left\{ -\alpha \pi(1+2N)
\right.\nonumber\\
& &+{\alpha \cos(\alpha)-\sin(\alpha) \over
\alpha} -{2N\cos(2N\alpha) -
\sin(2N\alpha) \over \alpha} \nonumber\\
& &+ {(1+2N)\alpha\cos[(1+2N)\alpha]
-\sin[(1+2N)\alpha] \over \alpha} \nonumber\\
& &+\left[2{\rm Ci}(\alpha)+\alpha\, {\rm Si}(\alpha)\right]
-(2N)\left[2{\rm Ci}(2N\alpha)+
2N\alpha{\rm Si}(2N\alpha)\right]\nonumber\\
& &+(1+2N)\left[2{\rm Ci}[(1+2N)\alpha]+
(1+2N)\alpha\,{\rm Si}[(1+2N)\alpha)]\right\}\nonumber\\
& &+ {1\over 2\tau} \int_{0}^{\infty}
dx\,L(x)\cosh^{-2}({x/ 2 \tau}),
\label{singimp}\end{eqnarray}
where $\tau$ appears in Eq.(\ref{tau}),
\begin{eqnarray}
& &\alpha=2\Lambda\Omega_d={2 \over \pi}
{a \over a_c}{T_d \over T_c},\nonumber\\
& &L(x)=\gamma_e+\ln(x)-\frac{1}{2}\left[
2N\ln(2N)-(1+2N)\ln(1+2N)\right]\nonumber\\
& &\quad-\frac{1}{4}\left\{
G(x)-2NG(2Nx)+(1+2N)G[(1+2N)x]\right\},
\end{eqnarray}
$G(x)$ is given in Eq.(\ref{hofx}), and the Cosine and Sine
integrals ${\rm Ci}(x)$ and ${\rm Si}(x)$ appear in
Eq.(\ref{sincos}). In Eq.(\ref{singimp}) we then take $a\to 0$ and
$N\to\infty$ in such a way as to keep the product
$2A=(2N+1)2a$ fixed. We first consider the zero temperature case,
by which this limiting procedure will yield the
critical thickness $A_c$ in this model below which
superconductivity cannot be maintained at any temperature.
We find
\begin{equation}
\frac{A_c}{a_c}=\frac{2}{\pi}\frac{T_d}{T_c}
e^{\gamma_e-1}=\exp\left[\frac{1}{gN(0)}-1\right].
\label{enhanced}\end{equation}
As for BCS superconductors $gN(0)$ ranges from about $0.2$
to $0.4$, this critical thickness $A_c$ is larger by about
a factor of 4 to 50 compared to the critical thickness $a_c$
obtained by the method described in Sec.~II, where the
pair potential was assumed to be constant over the entire
superconducting layer. As well, Eq.(\ref{enhanced})
reinforces the trend that the critical
thickness would be larger for weaker BCS superconductors with
smaller values of $gN(0)$, as one might expect.
\par
As with the case of thin
normal and superconducting layers of the finite lattice
in the last section, these features of the critical thickness
can be interpreted
in terms of the form of the pair potential suggested by Fig.~12,
where the gap parameters decrease relative to the central value
as one moves towards the edges. From this one might infer that
the superconductivity is in a sense ``weaker'' towards the
edges for such a potential compared to the case of a constant
gap parameter assumed over the entire superconducting region,
as in Fig.~1. If one pictures the breaking
of superconductivity as the ``leaking'' of Cooper pairs into
the normal region, then one would expect that the form of the
more ``realistic'' pair potential in Fig.~14 would lead to an
earlier destruction of superconductivity
compared to the constant pair potential of Fig.~1.
\par
One can also analyze the gap equation (\ref{singimp}) at finite
temperature. For example, in the vicinity of zero temperature
one finds
\begin{equation}
\frac{T_c(A)}{T_c}=\frac{\pi}{4}\frac{T_c}{T_d}
e^{1-\gamma_e}\left[\frac{A}{A_c}-1\right].
\end{equation}
However, at higher temperatures this method of analyzing the
single layer case fails, and in particular gives the wrong
bulk limit as $A\to\infty$. This would
indicate the assumption in this model of
a constant pair amplitude for thick initial superconducting
film layers is invalid.
\section{Conclusions}
We have considered the question of when a finite sized system
composed of alternating layers of superconducting and normal
metals can effectively be thought of as a true infinite superlattice.
Concentrating on the calculation of the critical temperature,
we extended some methods used previously for a single superconducting
layer and an infinite superlattice to such finite--sized systems.
By examining the form of the pair potential found
we argued that the approximation of a constant Cooper pair potential
in each superconducting region
seems reasonable for  thick superconducting films ($a/a_c>10$)
over a wide range of thicknesses of the normal films or else for
thin superconducting ($a/a_c<10$) and thin normal ($b/a_c<0.1$) films.
The results suggest that for thick superconducting films
it takes a relatively small number of layers ($N\sim 10$)
using these methods to effectively make
an infinite superlattice. However, for thin superconducting and
normal films noticeable differences were found in this
approach between the
finite and infinite superlattices, even for a relatively large
number of layers for the finite lattice. The corresponding decrease in
the critical temperature of the finite lattice at a fixed
thickness of the superconducting layer was attributed to the ``weaker''
superconductivity in the outer regions compared to the case of the
constant pair potential assumption of the infinite superlattice.
One might argue that such ``weaker'' zones in finite but large systems
would be true for more realistic forms of the pair amplitude
incorporating some spatial variation. We also applied these methods in
the case of a single superconducting layer to construct a model which attempts
to incorporate a more realistic form of the pair potential
than the assumed constant. The model suggested that
the zero temperature critical thickness will be
 enhanced with such a pair potential
compared to that found by assuming a constant pair potential, and
also exhibited the trend that weaker BCS superconductors will
have a larger value of this thickness. These features were
again attributed to the ``weaker''
superconductivity near the edges of a more realistic pair potential.
\section{Acknowledgements}
We thank Dr.~J.~P.~Whitehead for valuable discussions. RK thanks
the Institute for Theoretical Physics at Santa Barbara, CA, where
part of this work was done. This work
was supported by the Natural Sciences and Engineering Research
Council of Canada and by the National Science Foundation under
Grant No.~PHY89-04035.

\clearpage
\centerline{ {\bf Figure Captions} }
\begin{itemize}
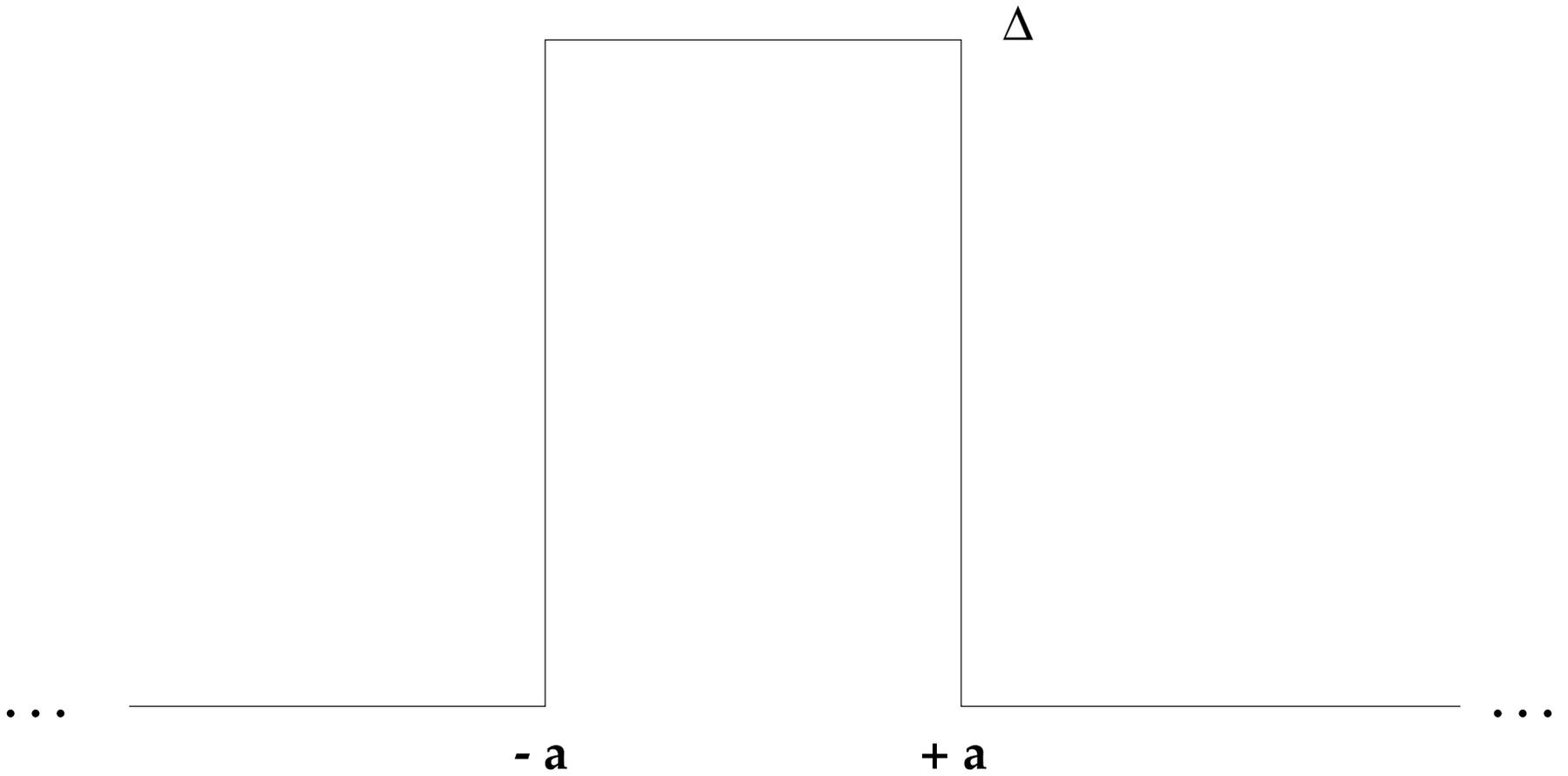
\item[Fig.~1:] The assumed form of the pair potential for a
single superconducting layer.
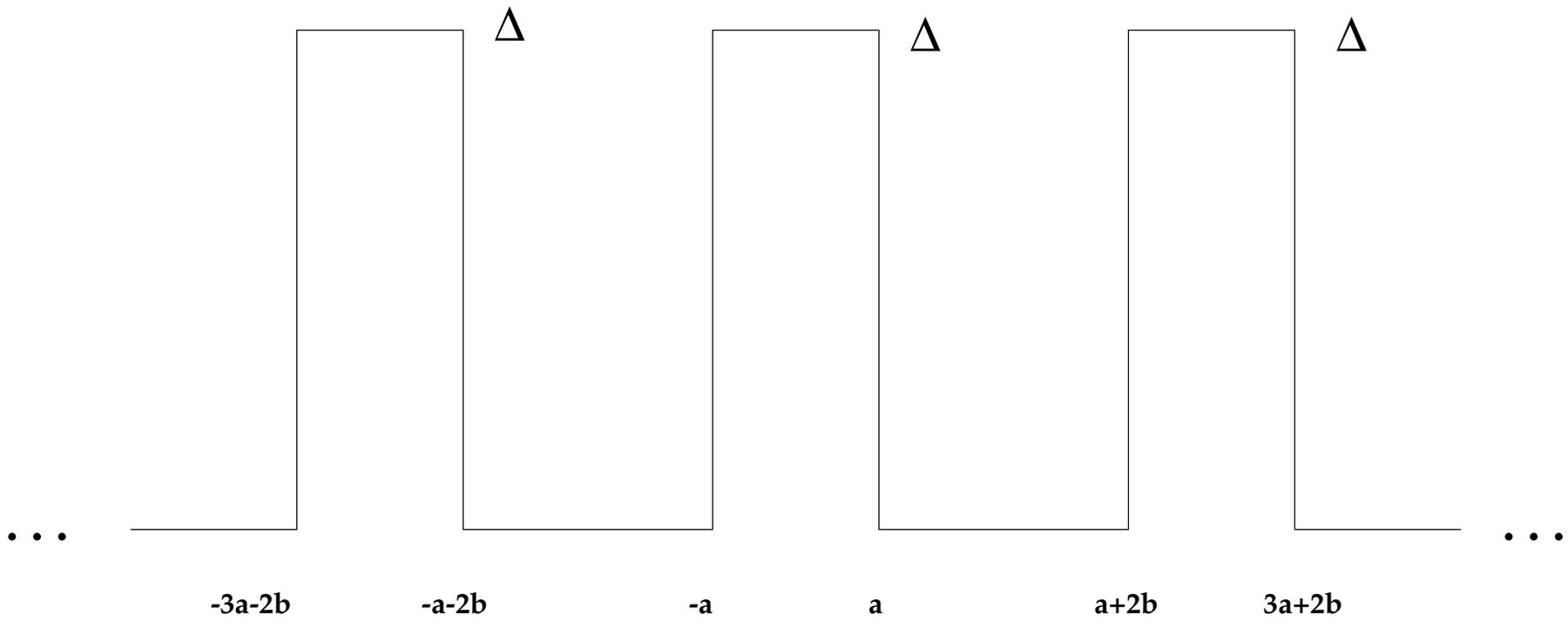
\item[Fig.~2:] The assumed form of the pair potential for an
infinite superlattice.
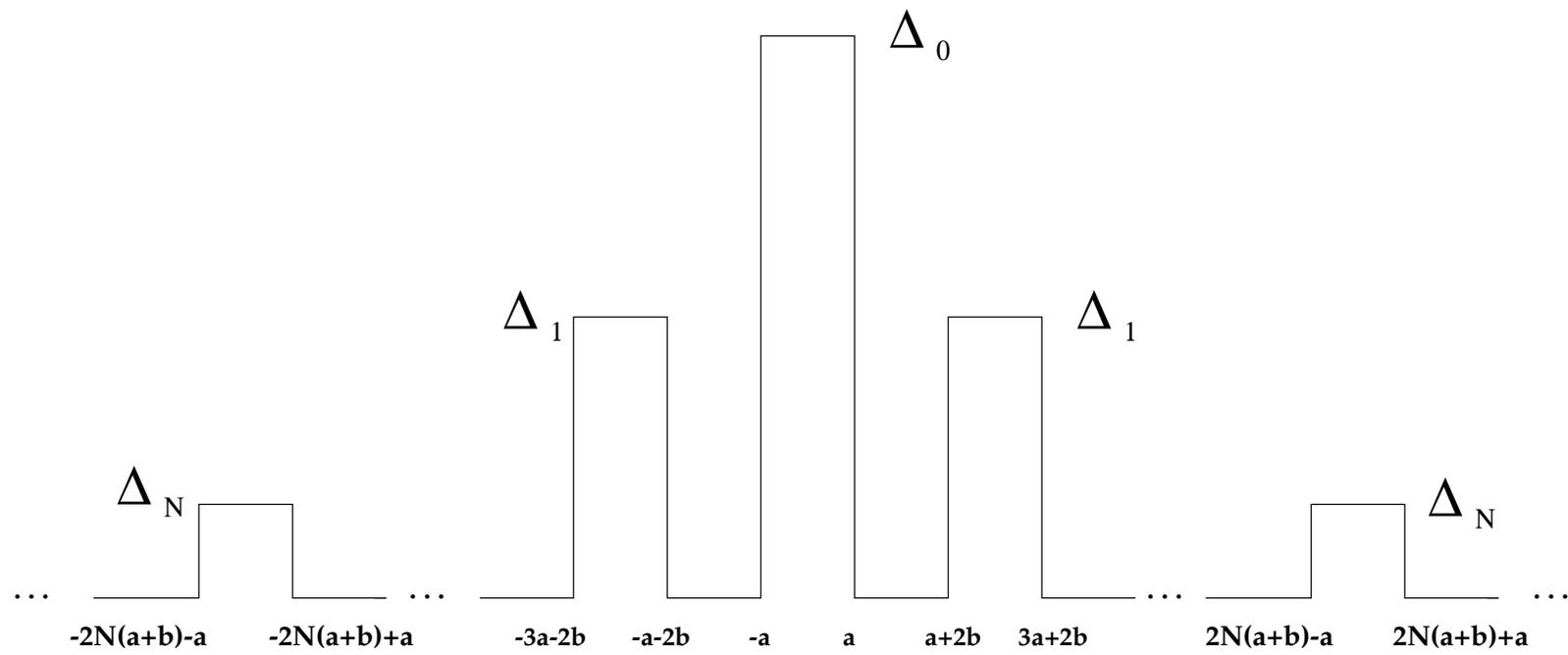
\item[Fig.~3:] The assumed form of the pair potential for a
finite ``superlattice'' of $2N+1$ layers.
\item[Fig.~4:] The critical temperature as a function of
$a/a_c$ for a single
layer, an infinite superlattice of $b/a_c=0.05$,
and a finite ``superlattice'' with $b/a_c=0.05$ and
$N=10$, $40$ and $100$.
\item[Fig.~5:] The critical temperature as a function of
$a/a_c$ for a single
layer, an infinite superlattice of $b/a_c=5.0$,
and a finite ``superlattice'' with $b/a_c=5.0$ and
$N=10$, $40$ and $100$.
\item[Fig.~6:] The gap parameter ratio $\Delta_j/\Delta_0$ for the
finite ``superlattice'' as a function
of the $j^{th}$ layer at zero temperature, with $N=40$ and $b/a_c=0.01$.
\item[Fig.~7:] The gap parameter ratio $\Delta_j/\Delta_0$ for the
finite ``superlattice'' as a function
of the $j^{th}$ layer at zero temperature, with $N=10$ and $b/a_c=0.01$.
\item[Fig.~8:] The gap parameter ratio $\Delta_j/\Delta_0$ for the
finite ``superlattice'' as a function
of the $j^{th}$ layer at zero temperature, with $N=40$ and $b/a_c=0.5$.
\item[Fig.~9:] The gap parameter ratio $\Delta_j/\Delta_0$ for the
finite ``superlattice'' as a function
of the $j^{th}$ layer at zero temperature, with $N=10$ and $b/a_c=0.5$.
\item[Fig.~10:] The gap parameter ratio $\Delta_j/\Delta_0$ for the
finite ``superlattice'' as a function
of the $j^{th}$ layer at finite temperature
$T_c(a,b,N)/T_c=0.283$, with $N=40$ and $b/a_c=0.5$.
\item[Fig.~11:] The gap parameter ratio $\Delta_j/\Delta_0$ for the
finite ``superlattice'' as a function
of the $j^{th}$ layer at finite temperature
$T_c(a,b,N)/T_c=0.282$, with $N=10$ and $b/a_c=0.5$.
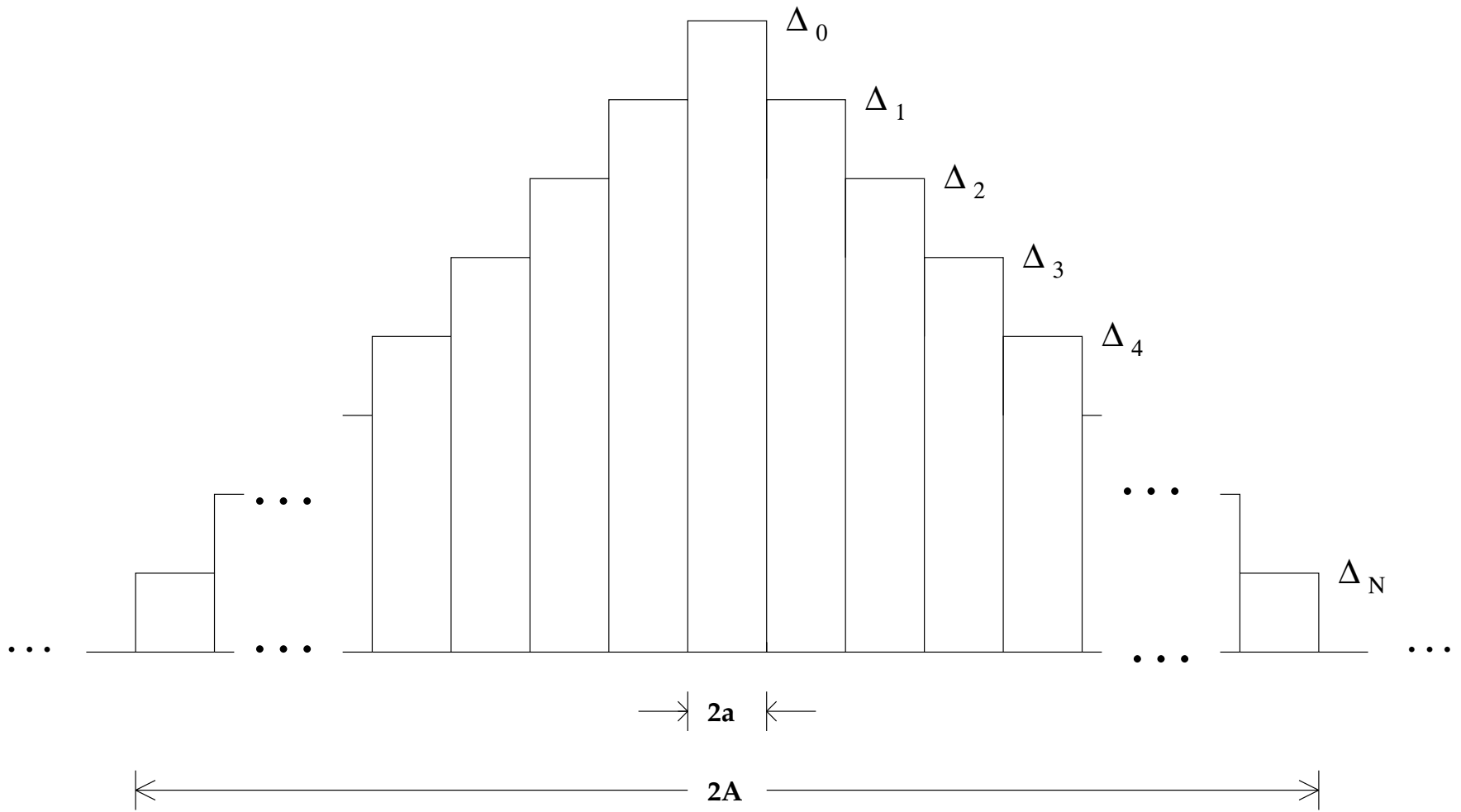
\item[Fig.~12:] The assumed form of the pair potential for the
single layer superconducting film of width $2A$ artificially divided into
$2N+1$ layers.
\end{itemize}
\end{document}